# Fast- and thermal-neutron detection with common NaI(Tl) detectors

Guntram Pausch[a], Achim Kreuels[a], Falko Scherwinski[a], Yong Kong[a], Mathias Kuester[a], Ralf Lentering[a], Andreas Wolf[a], and Juergen Stein*[a]

[a]Target Systemelektronik GmbH & CO. KG, Heinz-Fangman-Str. 4, D-42287 Wuppertal, Germany

## ABSTRACT

Radionuclide Identification Devices (RIDs) or Backpack Radiation Detection Systems (BRDs) are often equipped with NaI(Tl) detectors. We demonstrate that such instruments could be provided with reasonable thermal- and fast-neutron sensitivity by means of an improved and sophisticated processing of the digitized detector signals: Fast neutrons produce nuclear recoils in the scintillation crystal. Corresponding signals are detectible and can be distinguished from that of electronic interactions by pulse-shape discrimination (PSD) techniques as used in experiments searching for weakly interacting massive particles (WIMPs). Thermal neutrons are often captured in iodine nuclei of the scintillator. The gamma-ray cascades following such captures comprise a sum energy of almost 7 MeV, and some of them involve isomeric states leading to delayed gamma emissions. Both features can be used to distinguish corresponding detector signals from responses to ambient gamma radiation. The experimental proof was adduced by offline analyses of pulse records taken with a commercial RID. An implementation of such techniques in commercial RIDs is feasible.

**Keywords:** Scintillator, neutron detector, gamma detector, gamma spectroscopy, digital signal processing

## 1. INTRODUCTION

Gamma-ray spectroscopy and neutron detection are means to distinguish common (i.e., natural, medical, or industrial) sources of nuclear radiation from Special Nuclear Materials (SNM) comprised in nuclear reactors or weapons. Institutions serving homeland security or border and civil protection have therefore been equipped with standardized commercial instruments as Radionuclide Identification Devices (RIDs) [1] or Backpack Radiation Detection Systems (BRDs) [2] that allow searching for and classifying of radioactive materials.

Most RIDs and BRDs in use are based on detectors comprising a single NaI(Tl) crystal and a photomultiplier tube (PMT) catching the scintillation light. Such detectors are widely available, unpretentious with respect to construction, signal extraction, and signal processing, and provide good energy resolution at low cost. Advanced RIDs and BRDs providing an (optional) neutron sensitivity must either be equipped with an extra neutron sensor, or with a combined neutron-gamma detector featuring one of the cutting-edge scintillation materials developed in the last decades, as CLYC [3]-[4], CLLB [5], or NaIL™ [6]. This means, of course, an extra hardware expense which is reflected in the instruments' cost.

This paper demonstrates fast- and thermal-neutron detection with a commercial RID comprising a NaI(Tl) detector and electronics for digital signal acquisition and processing, but no dedicated neutron sensor. The neutron sensitivity and discrimination from gamma rays is obtained without any hardware modification, just by extending and refining the computational processing of digitized detector signals. Demonstrated offline, the corresponding algorithms shall be integrated in the instruments' firmware. Firmware-enabled neutron detection would increase the usability of common NaI(Tl)-based RIDs and BRDs.

*Juergen.Stein@target-sg.com; phone +49 202 76930 20; target-sg.com

## 2. PHYSICAL BACKGROUND AND PREVIOUS WORK

Neutron detection with NaI(Tl) is not a new idea. The crystal comprises iodine as a main constituent. Its only stable isotope, $^{127}$I, is distinguished by a quite large neutron-capture cross-section. The gamma rays following neutron capture in $^{127}$I have been exploited for fast- and slow-neutron detection with NaI(Tl) crystals since the 1950ies [7]-[8]. The corresponding signature interferes, however, with that of gamma radiation from other sources. In homeland security applications, where low-level neutron flux variations must be detected in the presence of (potentially strong) ambient gamma radiation, a simple background subtraction is not an option. Two feasible discrimination techniques have been explored so far:

- The first method, proposed in 2009 and named Neutron Capture Detector (NCD) [9]-[10], considers a high energy deposition in the scintillation crystal as distinctive feature. Common radioactive sources rarely emit gamma quanta of an energy above 2.614 MeV. Signals due to energy depositions between ≈ 3 MeV and the neutron separation energy $S_n(^{128}$I$)$ of ≈ 7 MeV can therefore be attributed to neutron captures in $^{127}$I. The remaining minor background is essentially due to cosmic radiation. It can be quantified by recording detector signals far beyond $S_n$, and by using this high-energy range for background normalization [11]-[12].

- The second technique exploits the delayed emission of distinct neutron-capture gamma rays. Sakharov *et al.* observed that gamma-ray cascades following neutron capture in $^{127}$I often involve the 137.8 keV (4$^-$) state of $^{128}$I, which has a half-life of 845 ns and decays to ground state via sequential emission of multiple low-energy gamma rays or conversion electrons [13]. Those are almost certainly absorbed within the NaI crystal and induce a scintillation response reflecting the 137.8 keV sum energy. Delayed coincidences between a first pulse, induced by the prompt part of the cascade, and a second pulse of about 138 keV, caused by deexcitation of the long-lived state, indicate neutron captures in $^{127}$I. This was demonstrated by Yakushev *et al.* in 2017 [14].

Meanwhile, the delayed-coincidence method has been refined and complemented with a triple-coincidence technique [15]. Gamma-ray cascades from $^{127}$I $(n,\gamma)^{128}$I sometimes involve not only the mentioned but one more isomeric level, the 167.4 keV (6$^-$) state of $^{128}$I with 175 ns half-life, which feeds the 137.9 keV (4$^-$) level with a strongly converted 29.5 keV transition. Such cascades induce triple-pulse sequences with energy depositions around 30 keV and 138 keV for the second and the third signal component, respectively. The half-lives of these states, and, consequently, the delays between sub-pulses of a corresponding signal, are comparable with the light decay time of NaI(Tl). An algorithm decomposing pulse pileups (instead of rejecting them) turned out to be the key for using delayed double- and triple-coincidence events for thermal-neutron detection in the presence of ambient gamma radiation [15].

The ongoing search for weakly interacting massive particles (WIMPs) – potential and so far hypothetic constituents of dark matter – has triggered interesting research to discriminate small scintillation signals of nuclear recoils, produced by fast-neutron scattering on sodium or iodine nuclei in high-purity NaI(Tl) crystals, from those caused by gamma- or X-ray interactions, i.e., by low-energy electrons [16]-[17]. So far, the sophisticated techniques developed in this context have only been applied to identify nuclear recoils in WIMP detectors but not to detect fast neutrons in common NaI(Tl) crystals.

Other studies explored the usage of neutron-induced nuclear reactions, knocking off protons, deuterons, or alpha particles of sodium or iodine nuclei, for fast-neutron detection [18]. Signals produced by the charged reaction products can be distinguished from gamma-ray interactions by means of pulse-shape discrimination (PSD), but the corresponding neutron detection efficiency is quite small and limited to high neutron energies.

## 3. FAST- AND THERMAL-NEUTRON DETECTION WITH A NaI(Tl)-BASED RID

To begin with an essential note: All graphs presented in this chapter were extracted from a single, coherent, 100-minute data acquisition accomplished in March 2021 at a fission neutron source (run_0024), complemented with a corresponding 100-minute acquisition at a gamma radiation source but in the absence of neutrons (run_0025).

### 3.1 Equipment and methods

Both measurements were performed with an F500 by Target Systemelektronik [19], a commercial RID licensed to Teledyne-FLIR and offered as identiFINDER®R440 [20]. It comprised an off-the-shelf scintillation detector with $\varnothing$2" × 2" NaI(Tl) crystal and Hamamatsu R10601 photomultiplier tube (PMT). The F500 electronics samples the PMT anode current directly with 14 bit resolution at a rate of 250 MS/s. The F500 was set up according to the product standard but operated with a modified firmware that allowed raw-data transfer of sampled detector signals to a PC. The raw data were stored on a file server in consecutive event records. For run_0024, the F500 was exposed to a $^{252}$Cf neutron source comprised in a polyethylene enclosure. For run_0025, the F500 was removed from the neutron source and exposed to gamma radiation of a sample comprising $^{232}$Th and its decay products. The source-detector distance was chosen such that the instrument's count rate equaled that of run_0024, namely 1.1 kcps. With the detector load in the target range, raw-data collection was started. Short calibration runs with a $^{137}$Cs source, performed before and after these measurements, complemented the campaign.

The data analysis was performed offline with Python scripts. After correcting for baseline offsets, the total and tail charges $q_{tot}$ and $q_{tail}$ were obtained for each pulse by integrating the PMT current signal over time in ranges from -12 to 3536 ns and from 168 to 3536 ns, respectively, where $t = 0$ corresponds to the constant fraction trigger point given by the F500 firmware. The total pulse charge was used to compute the energy deposition, assuming a simple linear dependence $E = c \cdot q_{tot}$. The calibration factor $c$ was extracted from calibration runs comprising the 662 keV gamma line of $^{137}$Cs as reference. Furthermore, all pulse records were searched for piled-up signal components. The stripping procedure deployed for pileup detection and decomposition provides separate energies and trigger points for all detected sub-pulses. It is described in reference [15].

### 3.2 Phenomenology of neutron effects in the PSD plot

Figure 1 presents plots of the tail-to-total charge ratio, a parameter often used for PSD, versus the energy deposition. The PSD plot measured at the neutron source (middle panel) exhibits several eye-catching structures that disappear in the absence of neutrons (left panel): a pronounced arc (1) above the main gamma ridge (2) which differs in shape from structures caused by random pileups (5), an enhanced branch of gamma rays with energies above 3 MeV (2), a bulge in the low-energy range at lower tail-to-charge ratio (3), and a sparse cloud of events (4) indicating knocked-out protons, deuterons, or alpha particles. Zooming in the lower-energy range (right panel) uncovers distinct substructures in the 'neutron' arc (1).

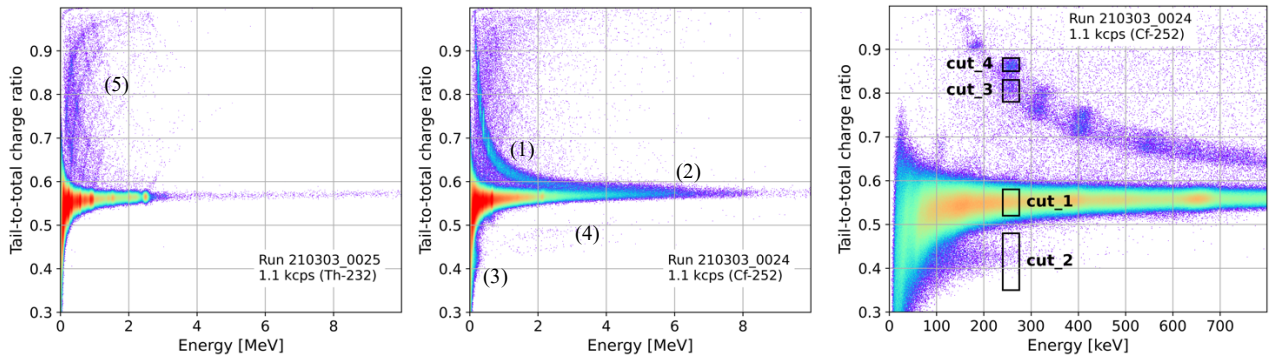

Figure 1. Standard PSD plot measured without (left panel) and with neutron source (center panel). The right panel zooms in the lower-energy part of the plot measured with neutron source, and exhibits the cuts used for the pulse shape inspection.

To identify the origin of the structures, we defined cuts in the PSD plot (Figure 1, right panel) and had a look the corresponding pulse shapes. The results are presented in Figure 2. Events in cut_1 and cut_2 are distinguished by very similar individual pulse shapes, respectively, but clearly different average shapes (Figure 2, left panel). Compared with the gamma-ray detections dominating the count rate (cut_1), the pulses from cut_2 exhibit a sharper rise and a shorter decay. These shape differences correspond closely to those reported by Spinks *et al.* [17] for nuclear and electronic recoils. This suggests the bulge (3) is due to sodium (and maybe also iodine) nuclei kicked off by fast neutrons. In contrast, most events in cut_3 or cut_4 comprise double- or triple-pulse sequences, respectively, often with piled-up components. Such signals occur if de-excitation cascades of $^{128}$I involve one or two isomeric states leading to delayed gamma emissions.

### 3.3 Fast-neutron detection

Exploiting the nuclear recoils, the PSD plot reveals for measuring fast neutrons [21] is obvious. The discrimination achieved with the conventional charge-comparison technique is, however, not convincing. Therefore, we tried a parameter used in WIMP search experiments: the mean time of a pulse, computed from all samples $S_i$ of a given signal, taken at times $t_i$ with respect to the trigger point, and defined as $t_{mean} = \sum t_i S_i / \sum S_i$ [16]-[17]. Good results were obtained when the time range used for the moment computation was restricted to about 1 µs, starting from the pulse onset.

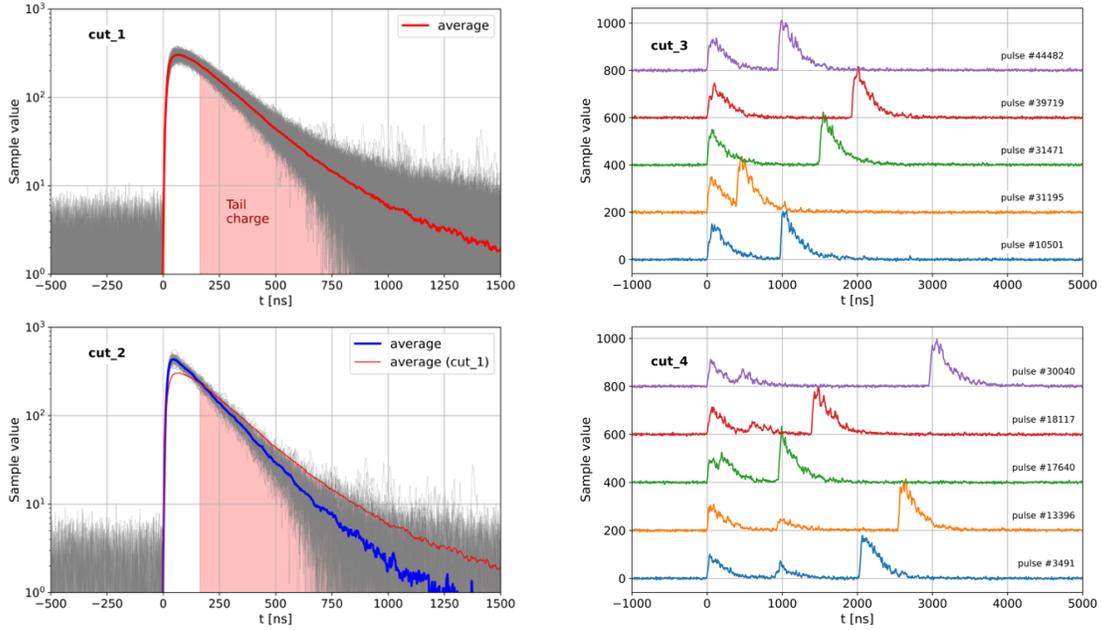

Figure 2. Left panel: Overlay of 200 individual pulses with the averaged signal for cut_1 and cut_2, respectively. The tail integration range is indicated. Right panel: Selected signals corresponding to cut_3 and cut_4.

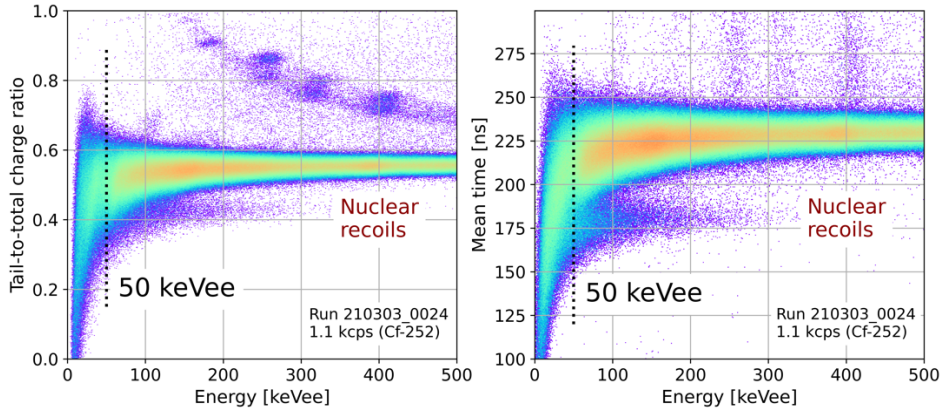

Figure 3. Comparison of tail-to-total charge ratio and mean time of a pulse as PSD parameters. The mean time provides for an improved separation of nuclear from electronic recoils down to a discrimination threshold of about 50 keVee.

Figure 3 compares a plot of $t_{mean}$ versus energy deposition with the standard PSD plot in the energy range of interest. The separation of nuclear recoils from gamma events is clearly improved. A discrimination threshold of 50 keV electron equivalent (keVee) seems to be feasible. With a quenching factor of 20-25% for Na recoils in this energy region [22], the threshold translates to 200-250 keV recoil energy. Such energies can be transferred with fast neutrons of 1.2-1.6 MeV. The energy deposition spectrum of nuclear recoils above the discrimination threshold should thus allow to distinguish fission neutrons from those produced in Pu-Be or Am-Be sources. Corresponding tests are envisaged.

### 3.4 Thermal-neutron detection

A comparison of energy deposition spectra measured with and without neutron source uncovers a significant fraction of events with energy depositions exceeding 3 MeV (Figure 4, left panel). They are essentially due to gamma-ray cascades originating from $^{127}$I neutron captures inside the crystal (where the chance of catching more than 3 MeV is increased as multiple gamma rays of a single cascade may interact with the crystal [10]), but could also comprise interactions due to energetic neutron-capture gamma rays produced in the crystal's environment. Corresponding events are selected with a cut involving the energy range from 3.2 MeV (which is in safe distance from the 2.614 MeV gamma line of $^{208}$Tl comprised in the $^{232}$Th decay chain) to 8.0 MeV (which includes most neutron separation energies). Data taken without neutron source exhibit a reasonably small background contribution. This background is essentially due to cosmic radiation, and it scales, consequently, with spectrum part above 10 MeV. Note the peak around 13-14 MeV is due to a controlled charge pulse clipping producing a compressed non-linear energy scale above 13 MeV.

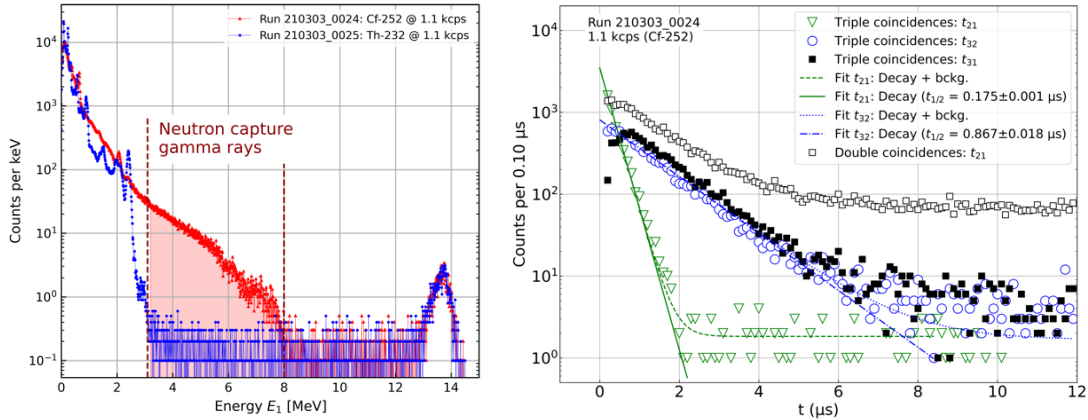

Figure 4. Left panel: Energy deposition spectrum (primary pulse component) with the cut selecting neutron-capture gamma rays. Background is due to cosmic radiation and scales with the spectrum part above 10 MeV. Right panel: Delay time distributions for double- and triple-pulse sequences. Fitted decay times reflect the lifetime of isomeric levels in $^{128}$I.

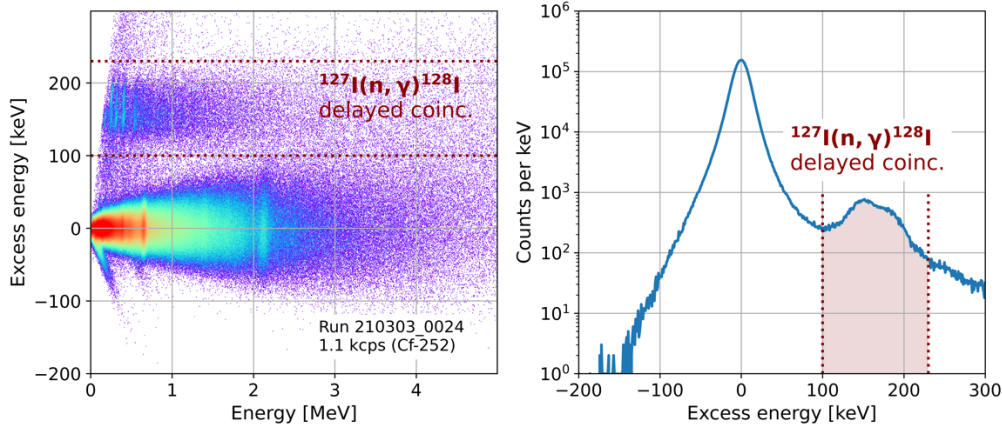

Figure 5. Excess energy comprised in the pulse tail, plotted versus total energy deposition (left panel). Excess energy or excess tail charge are well suited for separating delayed-coincidence signals from the bulk of single-pulse events before any (computationally expensive) pileup decomposition is applied. Right panel: Overall excess energy spectrum.

The right panel of Figure 4 presents delay-time distributions between sub-pulses of double- or triple-coincidence events as shown in the right panel of Figure 2. The delays were derived from the individual sub-pulse trigger times provided by the pileup decomposition algorithm [15]. The time distributions comprise exponential components with half-lives in correspondence with the lifetimes of the isomeric states in $^{128}$I discussed in Section 2. This confirms their interpretation as neutron-capture events in $^{127}$I involving one or both isomeric levels. Refence [15] explains in detail how such events could be selected and counted, supposed all pulse records were searched for delayed signal components and subsequently decomposed in their constituents. The latter is, however, computationally expensive and could, if implemented in the instruments' firmware and applied to all signals, reduce the achievable throughput.

The pileup decomposition should therefore be preceded with an efficient event pre-selection. As a matter of fact, a piled-up pulse sequence should, if compared with a signal consisting of the sole first sub-pulse, comprise an extra charge in the pulse tail in correspondence with the delayed energy deposition(s). We define a nominal tail charge $\langle q_{tail}\rangle$ as the average tail charge of single-pulse signals comprising the same lead charge as the signal considered, where lead charge means the charge $q_{tot} - q_{tail}$ comprised in the pulse excluding the tail. The excess energy $dE = c \cdot dq$, derived from the extra tail charge $dq = q_{tail} - \langle q_{tail}\rangle$, should then correspond to the delayed energy deposition(s), supposed the delayed signals are comprised in the tail of the primary pulse component. The left panel of Figure 5 presents the excess energy plotted versus total energy; the right panel shows the corresponding excess energy distribution. Single-pulse events should, per definition, not exhibit an excess energy; they are comprised in the dominating horizontal ridge (left panel) and in the main peak (right panel). Delayed double- and triple-coincidence events are distinguished by a delayed energy deposition of 138 keV or 30 keV plus 138 keV, respectively. Those cause the upper ridge in the 2D plot (left panel), and the shaded double peak in the excess energy spectrum (right panel). A cut in $dE$ (as sketched in Figure 5) or in $dq$ is obviously an appropriate pre-selector of potential delayed-coincidence events. Signals not comprised in the cuts need no pileup decomposition and could be treated with much less expense.

The delayed-coincidence ridge in the left panel of Figure 5 exhibits tiny substructures. It is worth noting that the narrow vertical stripes in the 0-1 MeV range reflect distinct feeding transitions of the isomeric levels in $^{128}$I. They allow to construct sophisticated, more selective event filters that could help to identify neutron-capture events even in intense ambient gamma-radiation fields. More details are given in [15] and [23].

## 4. CONCLUSIONS

A commercial RID, equipped with an off-the-shelf ∅2" × 2" NaI(Tl) scintillation detector, has been used to detect fast and thermal neutrons simultaneously and in parallel to ambient gamma radiation. Fast- as well as thermal-neutron signals could be separated from gamma-induced detector responses by applying multiple, tailored signal processing and pulse-shape analysis algorithms in parallel to directly sampled, digitized PMT anode signals. Neither the hardware itself nor any parameter of the instrument's setup (as PMT voltage, gain, threshold setting, shaping constant etc.) differed from that of an instrument leaving production. The only firmware modification concerned an option to transfer raw-data records (i.e., sampled detector signals) to an external file server.

Fast neutrons with energies above 1.2-1.6 MeV turned out to be detectible via nuclear recoils produced in the NaI crystal. PSD allowed their discrimination against gamma background for energies exceeding 50 keVee. This reasonably low detection threshold should allow to classify neutron sources by analyzing the measured recoil energy spectrum.

Slow neutrons could be assessed via gamma-ray detections following neutron capture in $^{127}$I. Those were distinguished from ambient gamma radiation by using two complementary methods: the detection of energy depositions in the 3-8 MeV range (NCD principle), combined with a background estimate based on the spectrum above 10 MeV, and the detection of delayed double- and triple-coincidences by pileup decomposition in combination with tailored cuts in the sub-pulses' delay time and energy parameters. Delayed-coincidence events are due to de-excitation cascades involving one or two isomeric states in $^{128}$I. The probability of such cascades is surprisingly high; corresponding estimates are given in [15].

State-of-the-art commercial instruments provide sufficient computing power to implement algorithms as discussed here in the instruments' firmware. RIDs or BRDs equipped with common NaI(Tl) detectors could thus be provided with reasonable neutron sensitivity at no extra cost, except for a firmware upgrade.